# Weyl Spinors with Proto-charges: A Substructure Model for Elementary Particles

**Hyung S. Choi**

Department of Mathematics, California Baptist University, Riverside, California 92504, U.S.A.

hchoi@calbaptist.edu

> "But there is another possibility that very few takes seriously: substructure – the idea that quarks and leptons (and maybe also the mediators) are composite particles, made of even more elementary constituents. This would change everything, just as the quark model changed everything 40 years ago, and Rutherford's atomic model changed everything a century ago. In any event, we almost certainly stand at the threshold of a fundamental revolution in elementary particle physics."
>
> --- David Griffiths, 2008 [1]

**Abstract**

We present a conceptual overview of a new substructure model of elementary particles, motivated in part by the Gell-Mann-Nishijima formula. In this approach, Weyl spinors endowed with proto-charges generate all known elementary fermions *via* their conjugate pairs, while bosons arise naturally as composites. The model predicts a new class of massive neutral bosons that serve as dark matter candidates, accounts for the existence of three generations of fermions, and reproduces the SU(3) x SU(2) x U(1) symmetry as a low-energy limit. Elementary particles acquire effective masses from internal binding energy, eliminating the need for *ad hoc* Yukawa couplings. Neutrinos emerge only in left-handed states, gain mass through their substructure rather than chirality mixing. From the simple proposition that elementary particles are Weyl spinors with proto-charges emerges a coherent and comprehensive picture of elementary particles and fields. Experimental implications of the model includes: no proton decay, absence of sterile neutrinos, absence of Majorana neutrinos, and a possible detection of new massive neutral bosons as dark matter.

**Keywords:** Gell-Mann-Nishijima formula; Weyl spinor; neutrino oscillation; dark matter; proto-charge; hexon.

## 1. Introduction

The Standard Model (SM) of particle physics is one of the most precise and comprehensive theories in the history of science. Developed in the 1970s on the SU(3)×SU(2)×U(1) gauge symmetry, it has withstood decades of experimental tests. Yet, new observations—most notably the evidence for dark matter and the discovery of neutrino oscillations—point to phenomena beyond its scope.

When a scientific theory reaches its mature stage, often small modifications risk disrupting its overall coherence. Extensions and *ad hoc* patches can accommodate new data, but at the cost of increasing complexity and diminishing elegance. History reminds us that the elaborate system of epicycles in the Ptolemaic model did not bring us closer to the truth [2].

Today's situation is reminiscent of earlier turning points. Before Rutherford introduced the nuclear model in 1911, the periodic table of elements suggested hidden order. Similarly, the "particle zoo" of the 1960s gave way to the quark model of Gell-Mann and Zweig, which explained hadrons as combinations of just three fundamental quarks. Time and again, mature has revealed her hierarchical structures. This invites us to ask: might elementary particles possess a deeper geometric and physical substructure? Could a simple principle underlie the diversity of particles we observe?

This paper offers such a case. We propose that fundamental particles are not point-like but composed of Weyl spinors carrying proto-charges. From this simple proposition emerges a coherent and comprehensive picture of elementary particles and fields.

## 2. The Substructure: Gell-Mann–Nishijima Formula

The Gell-Mann–Nishijima formula indicates that the electric charge of fermions arises from two distinct contributions— isospin and hypercharge. This decomposition hints at a possible substructure. Recasting these contributions in terms of a new fundamental unit, the proto-charge $\pm e_0$ where $e_0 = e/6$, we can represent all first-generation fermions in a uniform structural scheme. Here, the signs $+$and $-$ stand for $+e_0$and $-e_0$ respectively.

| | | |
|---|---|---|
| Electron: | $(- - -\vert - - -)$ | $T_3 = -\frac{1}{2}, \frac{Y}{2} = -\frac{1}{2}; \ Q = -1$ |
| Positron: | $(+ + +\vert + + +)$ | $T_3 = +\frac{1}{2}, \frac{Y}{2} = +\frac{1}{2}; \ Q = +1$ |
| Neutrino: | $(+ + +\vert - - -)$ | $T_3 = +\frac{1}{2}, \frac{Y}{2} = -\frac{1}{2}; Q = 0$ |
| Anti-neutrino: | $(- - -\vert + + +)$ | $T_3 = -\frac{1}{2}, \frac{Y}{2} = +\frac{1}{2}; Q = 0$ |
| 3 Up quarks: | $(+ + +\vert - + +), (+ + +\vert + - +), (+ + +\vert + + -)$ | $T_3 = +\frac{1}{2}, \frac{Y}{2} = +\frac{1}{6}; \ Q = +\frac{2}{3}$ |
| 3 Down quarks: | $(- - -\vert - + +), (- - -\vert + - +), (- - -\vert + + -)$ | $T_3 = -\frac{1}{2}, \frac{Y}{2} = +\frac{1}{6}; \ Q = -\frac{1}{3}$ |

where $T_3$ is the third component of the particle's Isospin and $Y$ is its hypercharge.

The substructure descriptions of quarks above present us with an interesting possibility that both electric charges and color charges may be combined into a single structure: leptons come in color-singlets and quarks comes in color triplets. As quarks come in three configurations, it is natural to identify them as the customary red, green, and blue:

| | | |
|---|---|---|
| $\vert -++)$ red, | $\vert +-+)$ green, | $\vert ++-)$ blue, |
| $\vert +--)$ anti-red, | $\vert -+-)$ anti-green, | $\vert --+)$ anti-blue, |
| $\vert +++)$ white, | $\vert ---)$ anti-white. | |

Each color charge has a different configuration of basic charges. Quarks and leptons thus emerge as different combinations of the same proto-charges. To avoid potential confusion, we introduce the following new terminology:

**Proto-charge:** a fundamental charge encompassing both electric and color properties.
**Triad:** a charge triplet $|\pm e_{01} \pm e_{02} \pm e_{03})$ representing one component of a spinor.
**Hexon:** a two-component Weyl spinor with proto-charges $(\pm e_{01} \pm e_{02} \pm e_{03} \,|\pm e_{04} \pm e_{05} \pm e_{06})$.

This construction unifies leptons and quarks at the substructural level, eliminating distinctions such as "chargeless" vs. "charged" or "colorless" vs. "colored."

The model requires two components and six degrees of freedom. Weyl spinor, the irreducible two-component spinor representation of $\mathrm{SL}(\mathbf{2}, \mathbb{C})$, offers a right framework. The algebra $\mathfrak{sl}(\mathbf{2}, \mathbb{C})$ consists of six generators (or six real degrees of freedom): three $J_i$ for rotations (space-space relationships) and three $K_i$ for boosts (time-space relationships). Defining two new generators:

$$J_i^{(+)} = \frac{1}{2}(J_i + iK_i) \tag{1}$$

$$J_i^{(-)} = \frac{1}{2}(J_i - iK_i) \tag{2}$$

which satisfy two independent SU(2) algebras: $\mathfrak{su}(\mathbf{2})_{L}$ and $\mathfrak{su}(\mathbf{2})_{R}$ where time-like transformations (boosts) are distributed between the two SU(2) groups. While a single (left- or right-handed) Weyl spinor *locally* has only three real degrees of freedom, their **bilinear** can encode the full six degrees of freedom of $\mathrm{SL}(\mathbf{2}, \mathbb{C})$. This is done in such a way that if you know one (say, a left-chiral spinor) then you know all the information about the other (its conjugate right-chiral spinor) while they are structurally different.

The electromagnetic field **F** also has three "boost-like" **E**-fields and three "rotation-like" **B**-fields. The complex combinations of them give us the self-dual and the anti-self-dual forms of the electromagnetic field that encode two helicities:

$$F^{(+)} = \vec{E} + i\vec{B} \qquad \text{Self-dual, corresponding to left-handed} \tag{3}$$

$$F^{(-)} = \vec{E} - i\vec{B} \qquad \text{Anti-self-dual, corresponding to right-handed} \tag{4}$$

In spacetime algebra (STA), the electromagnetic field is a single bivector field, not two separate vectors:

$$F = E^i \gamma_i \wedge \gamma_0(u) + IB^i \epsilon_{ijk} \gamma_j \wedge \gamma_k(u) \tag{5}$$

where $I = \gamma_0\gamma_1\gamma_2\gamma_3$. Hence, the same field *F* is viewed as different combinations of *E* and *B* depending on the observer's 4-velocity $u$. The spacetime structure represented by Weyl spinors is a native arena for electromagnetic fields. If we limit ourselves to the unit magnitude (quantum) of the field, each of these six degrees of freedom can have **two bivector directions** (positive and negative) in the plane of rotation – giving us exactly $2^6 = 64$ combinations.

**Spacetime from Weyl Spinors**

"Change" is one of the most immediate human experiences [3]. Time and space are our conceptual inventions to describe these changes. Élie Cartan saw spinors more fundamental than spacetime and thought that the very concept of rotation and orientation in space is dependent on the existence of spinors. Spacetime, as we classically understand it, emerges from interactions of spinors. [4].

A 2-component spinor defines a null vector plus an oriented 2-plane. This lets you build a time-like worldline. A parameter along the line can be interpreted as proper time. A system of interacting spinors can define relational time – i.e., how spinor A changes relative to spinor B. So starting from Weyl spinor, one can encode direction, phase, chirality, and causal structure – thus how spacetime arises from spinor geometry. The idea aligns deeply with Penrose's twistor theory, the loop quantum gravity, and other STA-based quantum gravity approaches. Some authors see this as one of the cleanest ways to rethink physics [5 – 6].

**Elementary Particles from Weyl Spinors**

Weyl spinors are often considered the building blocks of matter [5 – 8]. Irreducible representations of $\mathfrak{sl}(\mathbf{2}, \mathbb{C})$ may be classified by their spin values $(J_L, J_R)$ with dimension $n = (2J_L + 1)(2J_R + 1)$. Here $\left(\frac{1}{2}, 0\right)$ represents left-chiral Weyl spinor, and $\left(0, \frac{1}{2}\right)$ represents right-chiral Weyl spinor. Using the Clebsch-Gordon rule, various combinations of $\left(\frac{1}{2}, 0\right)$ and $\left(0, \frac{1}{2}\right)$ can generate (0,0), $\left(\frac{1}{2}, \frac{1}{2}\right)$, (1,0), (0,1), $\left(\frac{1}{2}, 0\right) \oplus \left(0, \frac{1}{2}\right)$ and so on, each of which listed here represents respectively: scalar, 4-vectors (potentially describing massive vector bosons), left-helicity photon, right-helicity photon, and the Dirac fermion. Simply put, Weyl spinors produce spins for all known fundamental particles in SM. With proto-charges, we have the complete combination of charges and spins supported by the spacetime structure.

## 3. The Hexon Model

We define a **hexon** as a Weyl spinor with proto-charges. The model admits $2^6 = 64$ basic spinors, corresponding to all possible proto-charge configurations. Table 1 shows the complete set of hexons in a symmetric manner. The table is isomorphic to the set of all possible **direct** products $Q_8 \otimes Q_8$ where $Q_8 = \{\pm 1, \pm i, \pm j, \pm k\}$, the quaternion group. This makes sense as any quaternion-valued field is equivalent to a SU(2) spinor structure *via* the corresponding relationships: $i \leftrightarrow i\sigma_1$, $j \leftrightarrow i\sigma_2$, $k \leftrightarrow i\sigma_3$.

For intuitive rendering, we used graphic symbols. The symbols here do **not** encode all the information about the system. It is a simple heuristic device that distinguishes the two chiral components and shows their proto-charge contents.

| Q ID C | w | r | g | b | $\bar{b}$ | $\bar{g}$ | $\bar{r}$ | $\bar{w}$ |
|---|---|---|---|---|---|---|---|---|
| w | +1 $e^+$ w | $+\frac{2}{3}$ u r | $+\frac{2}{3}$ u g | $+\frac{2}{3}$ u b | $+\frac{1}{3}$ $\bar{d}$ $\bar{b}$ | $+\frac{1}{3}$ $\bar{d}$ $\bar{g}$ | $+\frac{1}{3}$ $\bar{d}$ $\bar{r}$ | 0 $\nu$ w$\bar{w}$ |
| r | $+\frac{2}{3}$ u r | $+\frac{1}{3}$ rr | $+\frac{1}{3}$ rg | $+\frac{1}{3}$ rb | 0 r$\bar{b}$ | 0 r$\bar{g}$ | 0 r$\bar{r}$ | $-\frac{1}{3}$ d r |
| g | $+\frac{2}{3}$ u g | $+\frac{1}{3}$ gr | $+\frac{1}{3}$ gg | $+\frac{1}{3}$ gb | 0 g$\bar{b}$ | 0 g$\bar{g}$ | 0 g$\bar{r}$ | $-\frac{1}{3}$ d g |
| b | $+\frac{2}{3}$ u b | $+\frac{1}{3}$ br | $+\frac{1}{3}$ bg | $+\frac{1}{3}$ bb | 0 b$\bar{b}$ | 0 b$\bar{g}$ | 0 b$\bar{r}$ | $-\frac{1}{3}$ d b |
| $\bar{b}$ | $+\frac{1}{3}$ $\bar{d}$ $\bar{b}$ | 0 $\bar{b}$r | 0 $\bar{b}$g | 0 $\bar{b}$b | $-\frac{1}{3}$ $\bar{b}\bar{b}$ | $-\frac{1}{3}$ $\bar{b}\bar{g}$ | $-\frac{1}{3}$ $\bar{b}\bar{r}$ | $-\frac{2}{3}$ $\bar{u}$ $\bar{b}$ |
| $\bar{g}$ | $+\frac{1}{3}$ $\bar{d}$ $\bar{g}$ | 0 $\bar{g}$r | 0 $\bar{g}$g | 0 $\bar{g}$b | $-\frac{1}{3}$ $\bar{g}\bar{b}$ | $-\frac{1}{3}$ $\bar{g}\bar{g}$ | $-\frac{1}{3}$ $\bar{g}\bar{r}$ | $-\frac{2}{3}$ $\bar{u}$ $\bar{g}$ |
| $\bar{r}$ | $+\frac{1}{3}$ $\bar{d}$ $\bar{r}$ | 0 $\bar{r}$r | 0 $\bar{r}$g | 0 $\bar{r}$b | $-\frac{1}{3}$ $\bar{r}\bar{b}$ | $-\frac{1}{3}$ $\bar{r}\bar{g}$ | $-\frac{1}{3}$ $\bar{r}\bar{r}$ | $-\frac{2}{3}$ $\bar{u}$ $\bar{r}$ |
| $\bar{w}$ | 0 $\bar{\nu}$ $\bar{w}$w | $-\frac{1}{3}$ d r | $-\frac{1}{3}$ d g | $-\frac{1}{3}$ d b | $-\frac{2}{3}$ $\bar{u}$ $\bar{b}$ | $-\frac{2}{3}$ $\bar{u}$ $\bar{g}$ | $-\frac{2}{3}$ $\bar{u}$ $\bar{r}$ | -1 $e^-$ $\bar{w}$ |

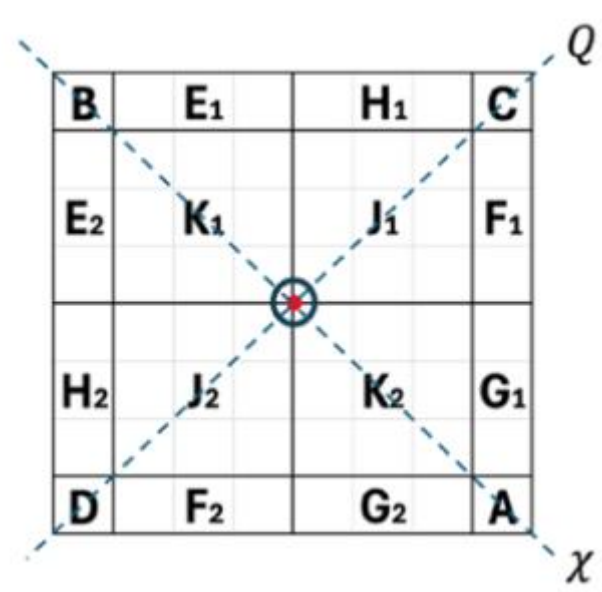

Table 1 Table 2

In Table 1, the first component of a Weyl spinor is represented by a left-pointing-triad, and the second component by a right-pointing-triad. Hence, we need both components together to form a hexon – a **single** Weyl spinor – which can be either left-chiral or right-chiral. We will find these graphical representations useful in tracking how particles are interacting with each other while preserving Lorentz transformation.

The table reproduces the known first-generation elementary fermions of in SM. Using the legend in Table 2, one can identify:

Electron (A) & Positron (B)
Neutrino (C) & Anti-neutrino (D)
Up quarks with three colors ($E_1$ left-chiral; $E_2$ right-chiral)
Down quarks with three colors ($F_1$ left-chiral; $F_2$ right-chiral)
Anti-up quarks ($G_1$, $G_2$) & Anti-down quarks ($H_1$, $H_2$)

The spinors above (A through H) represent the first generation of fermions and are located in outer sectors (A through H). As we will see, spinors in inner sectors $J_1$, $J_2$ and $K_1$, $K_2$ are combined to form spin-1 bosons.

In the legend located in the upper-left corner of the table, Q stands for electric charge; C for color charge; ID for particle identification such as electron, neutrino, up quarks, down quarks, and their antiparticles. The small letters in Table 1: w, r, g, b and $\overline{\mathrm{w}}, \overline{\mathrm{r}}, \overline{\mathrm{g}}, \overline{\mathrm{b}}$ represent the color and anti-color: white, red, green, blue and anti-white, anti-red, anti-green, and anti-blue, respectively. Left external column of the main table shows the first component of a Weyl spinor; Top external row shows the second component of a Weyl spinor.

The hexon diagram is arranged in such a way that a spinor that represents a particle is situated at the opposite location from its antiparticle's with respect to the center of the hexon table (see the center circle in Table 2). Note also that all particle-antiparticle pairs in the table have both the total charge and the total chirality zero.

There are other symmetries noticeable in the table: (1) The hexons that are located at the opposite positions about the dotted diagonal line Q in Table 2 have the same **total** charge with opposite signs. The hexons along the line Q – those in C, D, $J_1$, and $J_2$ – therefore have no net charges. (2) The the hexons located at the opposite positions about the dotted anti-diagonal line $\chi$ in Table 2 have the opposite chiralities. So, hexons along the line $\chi$ – those in A, B, $K_1$, and $K_2$ – have both left- and right-chiralities. Importantly, **neutrinos appear only with one chirality**, consistent with experimental observations, while other fermions retain both. Note that up quark has both chirality ($E_1$, $E_2$), so does the down quark ($F_1$, $F_2$).

Because of the charge substructure, the concept of antimatter in HM needs to be slightly revised. An antiparticle is still the charge-conjugate of a particle, but the idea of a charge-conjugate needs to include different charge configurations within a hexon. The simplest way to define this new charge conjugation is to flip each proto-charges. For instance, $C(++-|-+-) = (--+|+-+)$, etc. This will allow a color charge turns to its anti-color. In this particular example, both the particle and its antiparticle have a total charge zero, but they are not identical to each other.

This will have an important implication for neutrinos: in HM, a neutrino is not its own antiparticle as $C(+++|---) = (---|+++) \neq (+++|---)$. Thus **no Majorana particle exists in HM**. This single-chiral nature of neutrino comes directly from the fundamental symmetry of HM and it is unique among fermions.

**SU(3)×SU(2)×U(1) as Approximate Symmetry**

Note that each quadrant of the hexon table (Table 1) has a SU(4)-like structure which has SU(3) and U(1) as subgroups and decomposes into $8_0 + 3_{+1} + \overline{3}_{-1} + 1_0$. This is similar to the Pati-Salam SU(4) unification of quarks and leptons [9]. The key difference is that while the Pati-Salam model took the lepton number as the fourth color, in our model, we don't introduce another color – leptons are simply color-neutral. While the Pati-Salam model predicts new gauge bosons that converts quarks into leptons (thereby mediate proton decay), our model fundamentally distinguishes quarks and leptons *via* substructures – thus **no proton decay**. The first quadrant of Table 1 is shown as an example below in Table 3. Note that SU(3) in the first quarter gives us exactly the **gluon sector** in SM.

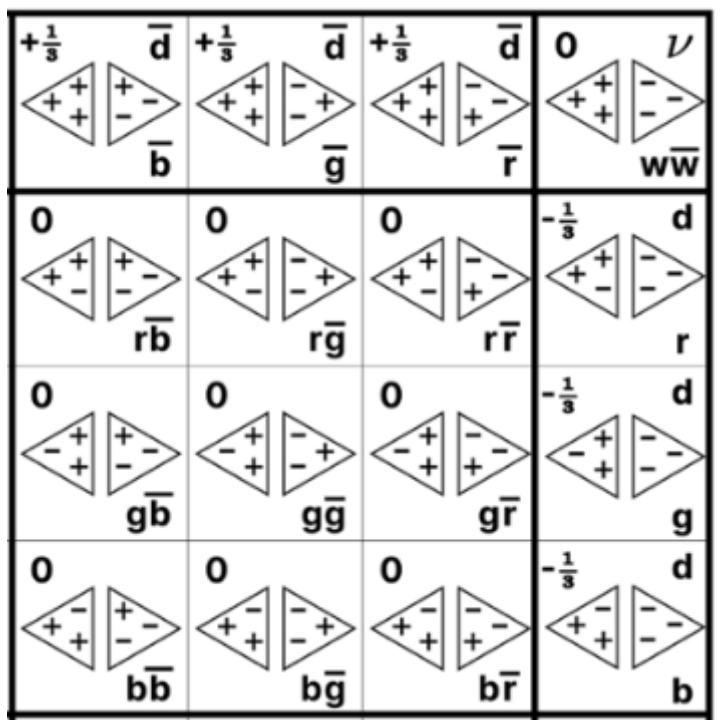

Table 3

The SU(4)-like structure in HM lives **within** the overall SU(2) that represents left or right Weyl spinors. We used the term "SU(4)-like" to distinguish our model with a substructure from the standard SU(4) group symmetry. SU(4) as a group, being compact 15-dimension, cannot sit inside $\mathrm{SL}(\mathbf{2}, \mathbb{C})$ that is noncompact 6-dimensions. In the energy scale that we have in our current particle accelerators, where we neither have enough energy to penetrate into the substructure with proto-charges nor the delicacy that would distinguish the different structural layers, the symmetry would appear much like SU(3) × U(1) × SU(2), where the first two symmetries come from SU(4)-like substructure.

**Dark Matter Candidate**

The only particles that are in the table but are not in the Standard Model are presented by the spinors in the sectors $K_1$ and $K_2$. (See Tables 1 and 2) Note that $K_1$ and $K_2$ must be combined to become color-neutral. As they are combined, they become electrically neutral as well. The simplest combination – for instance, the combination of $\langle gg \rangle$ and $\langle \bar{g}\bar{g} \rangle$, or $\langle br \rangle$ and $\langle \bar{b}\bar{r} \rangle$ – would be a massive neutral boson.

Three characteristics of this dark matter candidate in HM are clear: (1) As these neutral particles living in different sectors, they would not interact much with "normal" particles except for gravity; (2) As bosons, they would tend to be clumpy; and (3) As their compositions of proto-charges are more complex, they would be significantly heavier than usual matters. These features make them natural candidates for dark matter.

## 4. The Two-Component Structure of Weyl Spinors

In HM, particles interact and potentially change their identifies by exchanging their triads (components with proto-charges). Therefore, understanding the behavior of the two components of Weyl spinor is crucial. Consider a two-component spinor (say, left-handed):

$$\psi = \begin{pmatrix} \psi^1 \\ \psi^2 \end{pmatrix} \tag{6}$$

The two components mix under a Lorentz transformation:

$$\psi^{\alpha} \rightarrow M^{\alpha}{}_{\beta}\psi^{\beta}, \qquad M \in SL(\mathbf{2}, \mathbb{C}) \tag{7}$$

where *M* is a non-unitary complex $2 \times 2$ matrix.

$$M(\vec{\theta}, \vec{\eta}) = \exp\left(-\frac{i}{2}\vec{\theta} \cdot \vec{\sigma} - \frac{1}{2}\vec{\eta} \cdot \vec{\sigma}\right) \tag{8}$$

with a rotation by angle $\vec{\theta}$ and a boost by rapidity $\vec{\eta}$. In this general Lorentz transformation, the two components of a spinor mix and scale. Both in STA and in the Bloch sphere, the two components of a Weyl spinor may be considered homogeneous coordinates. Their relative magnitude fixes the polar angle – thereby sets the tilt of the spin/propagation direction; and their relative phase fixes the azimuthal angle – thereby sets the orientation of the transverse polarization plane.

However, when two spinors are interacting, one can choose the z-axis in the direction of their mutual motion (in the center of mass frame). Under rotation (for instance, a rotation by angle $\theta$ about the z-axis):

$$M_z(\theta) = \exp\left(-i\frac{\theta}{2}\sigma^3\right) = \begin{pmatrix} e^{-i\theta/2} & 0 \\ 0 & e^{i\theta/2} \end{pmatrix} \tag{9}$$

$$\psi^{\alpha} = \begin{pmatrix} \psi^1 \\ \psi^2 \end{pmatrix} \rightarrow \begin{pmatrix} \psi^1 e^{-i\theta/2} \\ \psi^2 e^{i\theta/2} \end{pmatrix} \tag{10}$$

Hence, $\psi^1$ acquires a phase of $-\theta/2$ while $\psi^2$ acquires a phase of $+\,\theta/2$. Likewise, under a boost by rapidity $\eta$ in *the z-direction:*

$$M_z(\eta) = exp\left(-\frac{\eta}{2}\sigma^3\right) = \begin{pmatrix} e^{-\eta/2} & 0 \\ 0 & e^{\eta/2} \end{pmatrix} \tag{11}$$

$$\psi^{\alpha} = \begin{pmatrix} \psi^1 \\ \psi^2 \end{pmatrix} \rightarrow \begin{pmatrix} \psi^1 e^{-\eta/2} \\ \psi^2 e^{\eta/2} \end{pmatrix} \tag{12}$$

Here, the first component is suppressed in positive boosts, the second component is enhanced. Thus, boosts change the relative importance of the two components. In massless limit, one component dominates to give us a helicity eigenstate. For massive particles, boosts can flip the helicity. The point here is that, in the direction of their interaction, two components don't mix – they just scale and change phases. This allows the exchange of spinors without destroying the structure of Lorentz transformation. See Section 6 for more spinor interactions.

**Topological Structure of Two-component Spinor**

The space of SU(2) is topologically represented by three sphere $S^3$. Each physical spinor corresponds to a point on $S^2$ (the Bloch sphere), and the fiber (circle $S^1$) corresponds to the phase associated with each spinor. A simple physics-motivated idea would be to consider $S^1$ as a directed loop with finite energy and angular momentum. This will give us dynamic stability and emergent particle-like properties. As a side note, Hopf fibration wouldn't work for our purpose because it lives in $\mathbb{R}^4$ rather than Lorentzian $SL(\mathbf{2}, \mathbb{C})$.

In condensed matter physics, we have physical analogs to this idea in quantum vortex rings in superfluid helium and Bose Einstein Condensates [10, 11]. In a superfluid, vortices are stabilized by quantized circulation and persistent flow with zero viscosity, and finite core energy and long-range velocity field – thus carry properties of a quantum of charge.

Once you have the charge arising from a such circulation (i.e., no delta-function-like source for charges), electricity and magnetism become fully symmetric and interconvertible under Lorentz symmetry. In this view, a particle is simply a localized field, and its energy and momentum are confined in the localized field – giving us an effective mass.

As an example, let's see the structurally simplest case of neutrinos. Note that the hexon diagrams are a simple rendering of Weyl spinors that live in 4-dimensional Lorentzian spacetime.

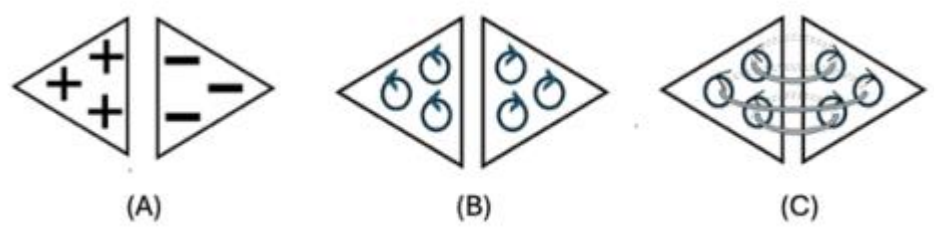

Figure 1

Each charge in (A) is considered equivalent to a circulating "magnetic" field. As we can see in (B), a positively directed loop acts like a source; a negative one acts like a sink. The figure (C) is the simplest way in which the sources and the sinks are linked.

This model integrates two different concepts in physics: source of a charge and a field generated by a charge. In this integrated view, everything is field. In this simple version, one can see that "**E**" and "**B**" fields are circulating around each other forming a bound state. This binding energy is identified as particle's effective mass. The concept is consistent with the fact that mass, charge, and a scalar from the field strengths $F_{\mu\nu}F^{\mu\nu}$ are all Lorentz invariants. In HM, they stem from the same substructure of the Weyl spinors.

**Three Generations of Fermions**

At the present stage of development, we do not know which specific geometrical or topological configurations (possibilities include flux tubes, helices, braids, rings, knots, twists, and vortices, etc.) will produce the observed mass spectra. This will be a rich area of study. However, the very structure of the hexon points to the fact that there are three levels of complication.

In HM, the three generations of fermions arise as different types of bound states stemming from the substructure without needing to externally introduce the family structure. A Weyl spinor has two components each of which encodes three internal degrees of freedom representing three color charges R, G, B. The labels are distinctive yet equally contributing to particle dynamics.

While there are many different topological or geometrical possibilities for connecting the two components of a spinor, in the end, the eventual "connection" between these two components can happen in only three qualitatively distinctive ways.

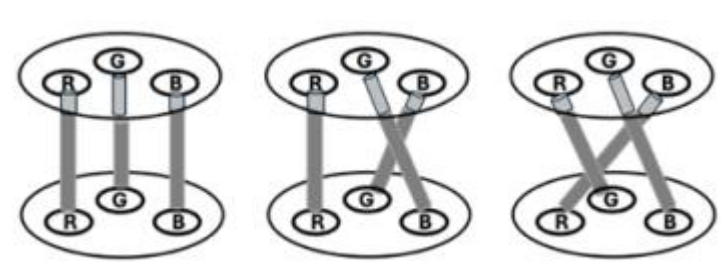

Figure 2

(Type I) Trivial pairing: (R-R, G-G, B-B). Match the same colors: It represents the simplest and lowest energy configuration – the first generation of fermions.

(Type II) One transposition: (R-R, G-B, B-G) or (R-G, G-R, B-B) or (R-B, G-G, B-R). Two components swap while one remains fixed. This represents the 2nd generation. These three variations of Type II all end up with the same mass as colors are only labels.

(Type III) All swap: (R-G, G-B, B-R), (R-B, G-R, B-G). This is the most complicated of the three configurations and may represent the 3rd generation.

**Two-component Spinor and Dirac's Belt Trick**

We have seen that spatial rotation and boost affect the phase and the rapidity of two components in opposite directions. If the two components are topologically connected as in our model, the spinor can get twisted quickly. However, the Dirac's belt trick provides us with a stable spinor behavior. Here we give a "free space" version – i.e., spinor is not attached to any external object.

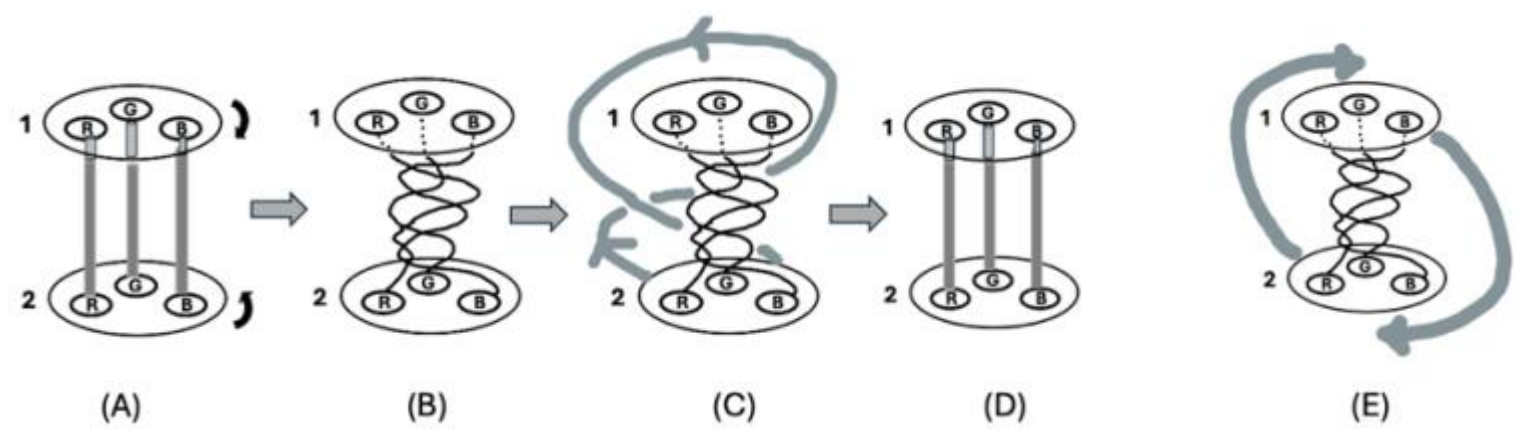

Figure 3

A $2\pi$ phase change of each of two components of a spinor in opposite directions amounts to a $4\pi$ internal twist. This (A) → (B) transformation is equivalent to twisting a belt by $720^0$ in Dirac's belt trick. The simple untangling mechanism in Dirac's original version is to move one end (the buckle-side) fully around the other side of the belt (the fixed end) without rotating the buckle itself. In a two-component spinor, it is the same as moving one component of the spinor – say, the component 2 in (C) – around the other component fully once without changing the color configuration (i.e., without rotating the triad itself). This will get back to the original configuration.

(E) shows that the process (C) is equivalent to moving the two components half-a-turn around each other in opposite directions without rotating the components themselves. In actual process, the tangling (A) → (B) and the untangling (E) occur simultaneously so that a spinor stays not tangled anytime. The untangling action (E) can be understood in an angular version of the hyperbolic rotation by defining $\sin\theta \equiv v/c$. Since $\tanh\eta = v/c$, we can map $\eta$ to $\theta$: when the rapidity $\eta$ varies from $-\infty$ to $+\infty$ (i.e., making a full hyperbolic rotation), the angle $\theta$ varies from $-\pi/2$ to $+\pi/2$ which amounts to a half-a-turn of a component of the spinor. Thus, as it is shown in (E), the untangling process requires each component of a spinor making a full hyperbolic rotation in opposite directions. Since the rapidity doesn't spatially rotate the components themselves, the color configurations of both components must be maintained during these hyperbolic rotations. Here one can see how a Weyl spinor stay untangled as each of the two components are "rotating" fully once in the opposite directions both in phase and in rapidity. It is also interesting to note that this triple helix can come in two ways – either left- or right-handed.

## 5. Vacuum and Mass Generation

We consider the physical vacuum as a self-contained sea of left- and right-chiral spinors. The idea that the vacuum is not a state of nothing, but a sea of spinors, is compatible with quantum field theory. In HM, like Higgs in SM, a scalar field is pervading spacetime but in the form of paired chiral spinors. As both vacuum and Higgs are in the state of total spin 0 with even parity $J^P = 0^+$, we simply consider Higgs an excited or condensate state of the vacuum.

**Mass Generation for Fermions**

In SM, a particle gains mass when its field couples with Higgs field *via* Yukawa coupling. There, Yukawa couplings are arbitrary constants that need to be put in by hand to produce the observed masses. In contrast, due to its substructure, HM offers a possibility for particles to get masses from its "first-principles."

Since HM has the same basic substructure for all fundamental particles, we can afford a simple and universal way for all fermions to acquire their masses. In HM, the physical vacuum is full of chiral spinors with proto-charges. Therefore the two types of interactions are possible in the vacuum: (A) **like-chiral interactions** (left-left or right-right as in photons in Section 6) or (B) **opposite-chiral interactions**. Like-chiral interactions don't generate mass, therefore result only in momentum propagation (e.g., light). Opposite-chiral interactions are involved in mass generation—just as in SM.

Fermion masses arise from two mechanisms: 1. **Effective mass:** binding energy between two triads; 2. **Proper mass** *via* a Lorentz scalar term in Lagrangian**:** A natural way for a fermion to get its mass is from the binding energy between a hexon (particle) and its C-conjugate (antiparticle).

All fermions may be created from vacuum in pairs with opposite-chiral spinors. Chirality-wise:

$$|\Psi\rangle_{vacuum} \;\rightarrow\; |\psi_\chi\rangle \otimes |\bar{\Psi}_{\bar{\chi}}\rangle \tag{13}$$

where $\chi$ and $\bar{\chi}$ represent opposite chiralities. Angular momentum-wise:

$$|0,0\rangle_{vacuum} \rightarrow \frac{1}{\sqrt{2}}\left(\left|\frac{1}{2},-\frac{1}{2}\right\rangle - \left|-\frac{1}{2},\frac{1}{2}\right\rangle\right)_{two\ entangled\ spinors} \tag{14}$$

The entangled nature of a particle-antiparticle pair from vacuum is crucial in understanding how the six degrees of freedom of SL($\mathbf{2},\mathbb{C}$) is distributed to $SU(2)_L$ and $SU(2)_R$. All conjugate hexons are "born" entangled and each carries the information of the whole just as one strain of DNA carries the information of the whole DNA.

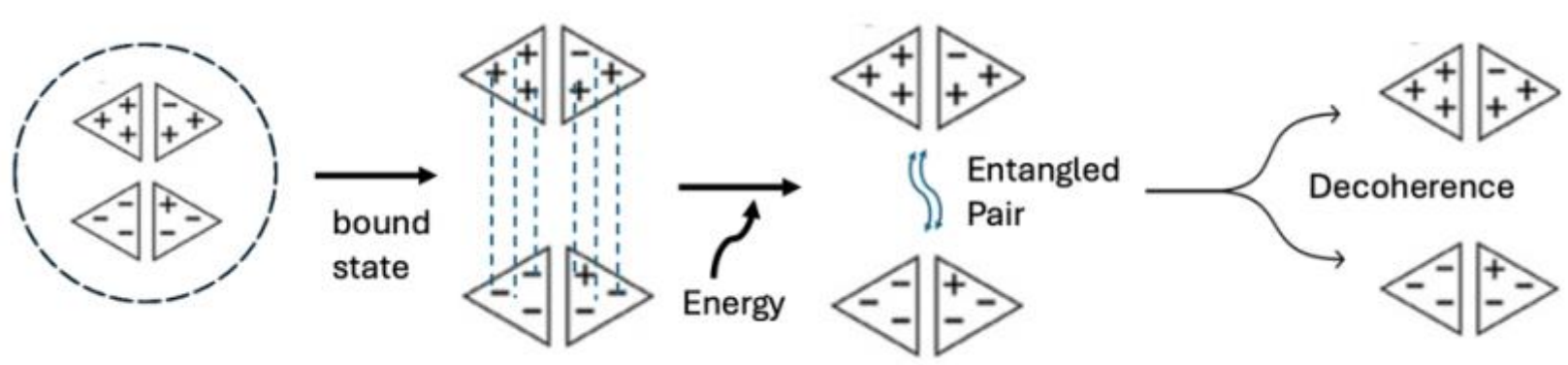

Figure 4

Pick a quark as an example (a red quark/anti-red quark pair as shown above). When energy is applied to a hexon and its conjugate, the binding between them due to opposite charges will be broken to produce an entangled pair. Note that all six proto-charges have opposite signs and the pair are bound by the forces due to their arrangements of proto-charges (both internal arrangement within a hexon and external ones between the hexon pair). The pair is entangled in all aspects due to the conservation of total chirality, total charges, total momentum, and total angular momentum. Except for neutrinos, fermions have net charges. Therefore they quickly decohere in normal circumstances and become independent particles.

The Dirac mass term in chiral basis:

$$m\bar{\psi}\psi = m\psi^\dagger\gamma^0\psi = m(\psi_L^*\psi_R + \psi_L^*\psi_R) \tag{15}$$

works as a Lorentz scalar for all fermions (except for neutrinos) as it mixes left- and right-chiral fields. In operator language, the $\bar{\psi}\psi$ term also mixes particle and antiparticle creation/annihilation operators, ensuring that the particle and its antiparticle have the same mass.

**Why Neutrino Mass is So Small**

For a neutrino, the second type of mass ("proper" mass) term exists only when it (left-chiral) is combined with its anti-neutrino (right-chiral) and lasts while their quantum coherence is maintained. Once coherence is disrupted, the neutrino can no longer access its right-chiral partner, and the corresponding mass term disappears through interactions with the environment. Other fermions, possessing both left- and right-chiralities, retain their mass terms under such conditions. Neutrinos, in contrast, having only one chirality, end up losing this contribution and are left only with a small effective mass arising from the binding energy between two triads. This explains their unusually small but nonzero masses.

## 6. Particle Interactions and Flavor Changes

In Sections 1 & 2, we have identified a triad (a single component of a Weyl spinor) as the common framework for color and electric charges. The binding within triads (the "color force") is stronger than the binding between the triads (the "weak force"). So, triads themselves are not altered in normal particle interactions. All known particles and the dark matter candidates (see Section 3) are different combinations of these two "unalterable" triads. As these triads are exchanged, particles change their identities or flavors. We identify these interactions that swap the triads as the weak interaction in SM.

As shown in Section 4, the two components of a Weyl spinor are not just placeholders – each component transforms differently under the spacetime symmetry group. Swapping these components, say $\psi^1 \leftrightarrow \psi^2$, is a change of the spinor itself and therefore amounts to a change of physics. Since we need to keep two components distinctive as they transform differently under Lorentz transformation, we can only exchange a first triad with another first triad, and a second triad with another second triad. This significantly limits possible interactions.

We thus place two rules for particle interactions: (1) no altercation of triads and (2) exchange of triads from the same component only. These rules preserve all conserved quantum numbers (electric charges, color charges, lepton number) and generally preserves approximate conserved quantities such as isospin. This works for all known interactions. Hence, unlike SM, we treat the weak interaction as a higher-level mechanism that enables the exchange of triad pairs, therefore often results in particle's change of identities.

A couple of examples are given below for illustration.

**(a) Neutron to proton beta decay: $d \rightarrow u + e^- + \bar{\nu}$**

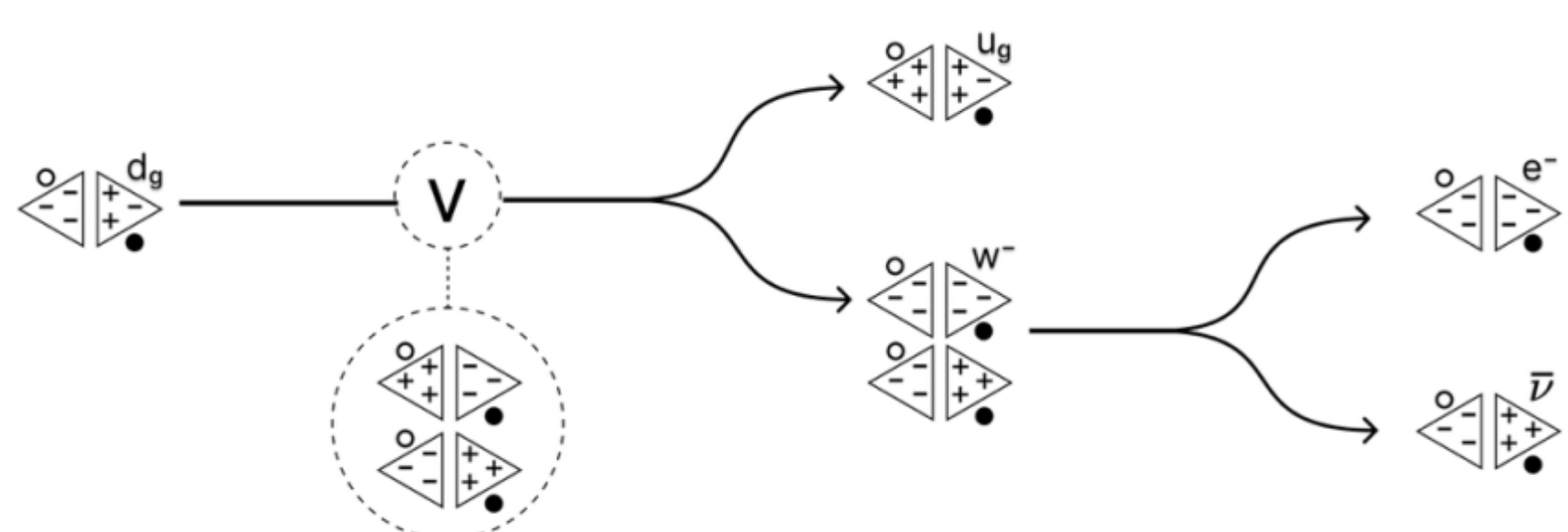

Figure 5

This example shows that particles exchange only the same components: the first component change places only with other first component (indicated by white dots) and the second component only with other second component (black dots). Here, a Higgs particle in SM may be interpreted as a vacuum excitation (represented here by V with dotted circle) that carries enough energy to produce both an up quark and a $W^-$ boson.

**(b)** $\boldsymbol{e^+ + e^- \to \nu + \bar{\nu}}$ **(via** $\boldsymbol{Z^0}$**)**

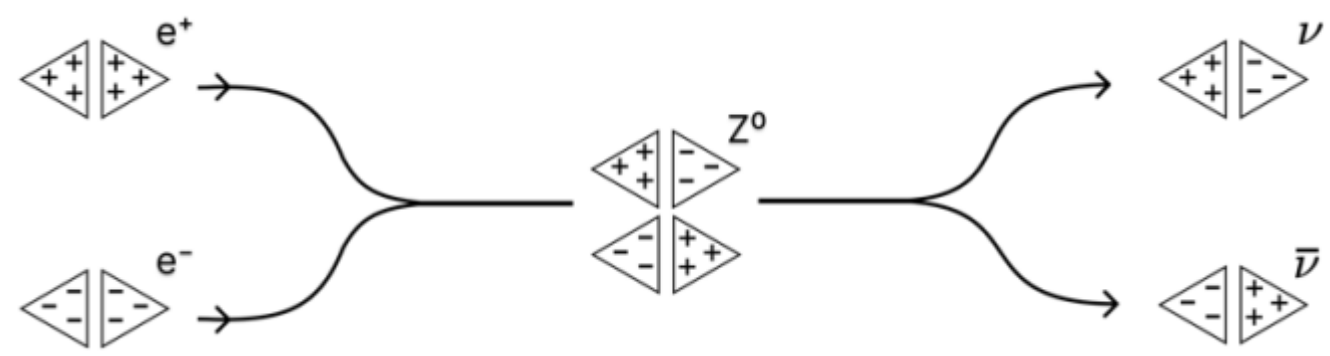

Figure 6

As in SM, two spinors don't simply produce two other spinors directly unless an additional structure, such as a scalar or a mediating field, is introduced. Here the spin-1 massive boson is mediating the process and is considered a composite of two Weyl spinors that **already have masses**. So, in HM, we don't need the Higgs mechanism for the mass of $Z^0$. In this example, the mixing of a positron and an electron yields $Z^0$which subsequently decays into a neutrino and an anti-neutrino.

**Photon as Interacting Weyl Spinors**

In HM, a photon is a composite of two massless Weyl spinors that **does not involve an exchange of triads** and thereby stay massless. Taking the symmetric tensor product of two fundamental left-handed Weyl spinors, we get: $(½, 0) \otimes (½, 0) = (1, 0) \oplus (0, 0)$. Here we used the $(j_L, j_R)$ labeling of Weyl spinors. Likewise, from the two right-handed spinors, we get $(0, ½) \otimes (0, ½) = (0, 1) \oplus (0, 0)$. As we are dealing with the antisymmetric tensor for massless bosons, the scalar part $(0, 0)$ doesn't contribute and simply returns to vacuum. So, we get the antisymmetric rank-2 tensor representation $(1, 0) \oplus (0,1)$ which corresponds to the field strength tensor $F^{\mu\nu}$.

$$F^{\mu\nu} = \psi_L^\dagger \sigma^{\mu\nu} \psi_L + \psi_R^\dagger \bar{\sigma}^{\mu\nu} \psi_R, \tag{16}$$

In terms of Weyl spinors, we rewrite (in the van der Werden spinor notation).:

$$F^{\mu\nu} = (\sigma^{\mu\nu})_{\alpha\beta} \psi_L^\alpha \psi_L^\beta + (\bar{\sigma}^{\mu\nu})_{\dot{\alpha}\dot{\beta}} \psi_R^{\dot{\alpha}} \psi_R^{\dot{\beta}}, \tag{17}$$

where $\sigma^{\mu\nu}$ and $\bar{\sigma}^{\mu\nu}$ are the generators of Lorentz transformations in the spinor representation.

**Pair Annihilation:** $\boldsymbol{e^+ + e^- \to 2\gamma}$

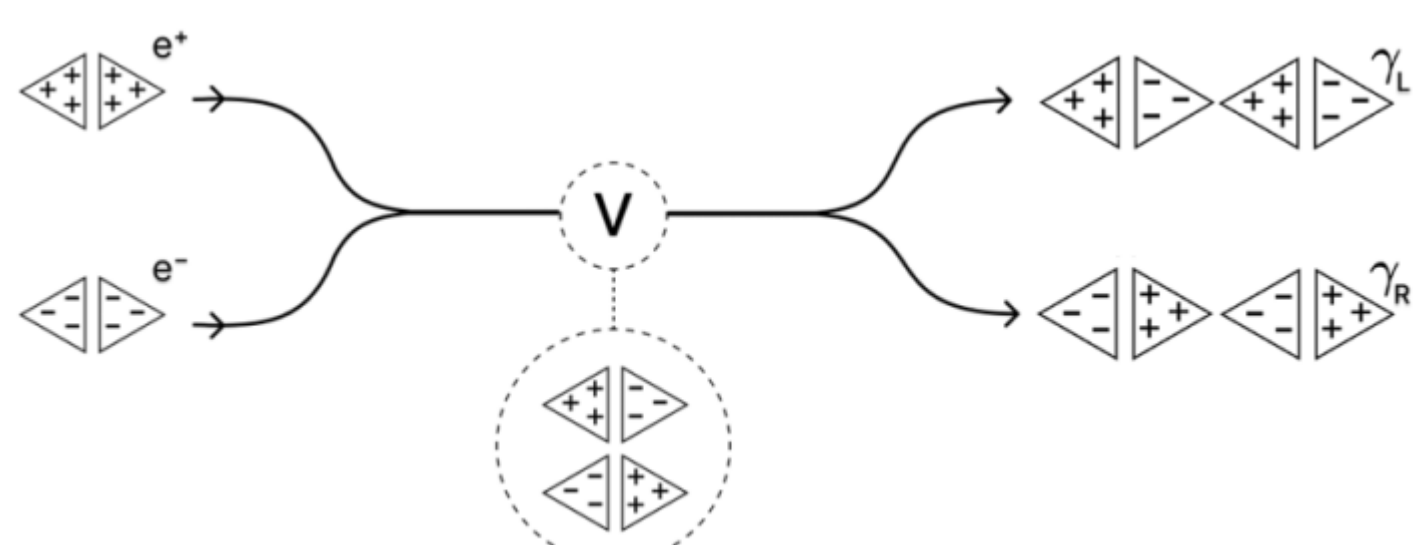

Figure 7

The massless fields can be described by a single Weyl spinor (left- or right-handed). And only one chiral spinor is needed to define a direction and polarization. Given a 2-component—say left-handed—Weyl spinor $\psi_L$, form the real null vector:

$$k^\mu = \psi_L^\dagger \bar{\sigma}^\mu \psi_L \tag{18}$$

where $\bar{\sigma}^\mu = (I, -\vec{\sigma})$.

Note that $k^\mu k_\mu = 0$. It points in the light-like direction and encoding the momentum of a massless particle. Note that in case of photons, two left spinors are graphically placed side-by-side emphasizing its charge configuration that is different from the cases that involve both chiralities. In this case, the field is not trapped but propagates in opposite directions. In the case of pair annihilation described above, in its center of mass reference frame, the opposite-chiral photon pair breaks away in opposite directions. So, they are observed to have the same helicity.

**Massive Bosons as Composites of Weyl Spinors**

As we have seen above, spin-1 bosons are considered the composite of two spin-1/2 Weyl spinors. A vector boson with $J = 1$ can arise as a composite of a left-handed and a right-handed Weyl spinor:

$$V^\mu \sim \psi^\dagger \sigma^\mu \chi. \tag{19}$$

This transforms as a Lorentz four-vector, which includes a massive vector boson: $(½, 0) \otimes (0, ½) = (½, ½)$ . This composite structure explains the mass of gauge bosons without needing to resort to a Higgs particle. One example of this is depicted in Figure 6 where the intermediate boson $Z^0$is formed from the spin-½ particle-antiparticle pair by exchanging their second spinor components.

Here we can have four massive bosons ($W^+, Z^0, W^-,$ and $H^0$) as a triplet and a singlet, $2 \otimes \bar{2} = 3 \oplus 1$, as we combine two 2-dimensional (spin-1/2) representations of SU(2).

$$\text{Triplet} \begin{cases} |1,1\rangle = \left|\frac{1}{2},\frac{1}{2}\right\rangle \\ |1,0\rangle = \frac{1}{\sqrt{2}}\left(\left|\frac{1}{2},-\frac{1}{2}\right\rangle + \left|-\frac{1}{2},\frac{1}{2}\right\rangle\right) \\ |1,-1\rangle = \left|-\frac{1}{2},-\frac{1}{2}\right\rangle \end{cases} \qquad \text{Singlet } |0,0\rangle = \frac{1}{\sqrt{2}}\left(\left|\frac{1}{2},-\frac{1}{2}\right\rangle - \left|-\frac{1}{2},\frac{1}{2}\right\rangle\right) \tag{20}$$

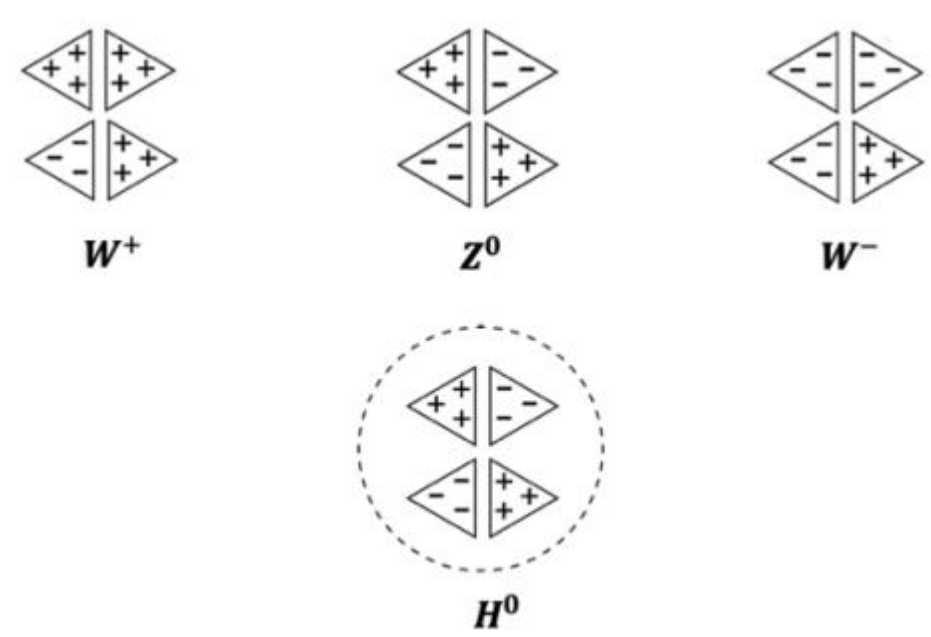

Figure 8

Higgs $H^0$ in HM is a singlet spin state $|s, m_s\rangle = |0, 0\rangle$ and naturally arises from the "featureless and symmetric" vacuum – combining one left-chiral spinor with one right-chiral spinor. **There is no hypothetical charged Higgs $H^+$**. Here, Higgs particle is presented with a dotted circle to distinguish it from a vector boson $Z^0$.

## 7. Neutrino Oscillation

HM provides a natural explanation for neutrino phenomena: (1) Neutrinos oscillate due to differences in internal binding energies rather than conventional mass states. (2) All neutrinos are left-handed, with no sterile or right-handed states. (3) Neutrinos are distinct from their antiparticles; thus, no Majorana neutrinos exist. (4) Thus, the neutrino-less double beta decay ($0\nu\beta\beta$) is forbidden. (5) Neutrinos may possess a small electric dipole moment (EDM), which would signal their effective mass and substructure. These predictions are consistent with current data and provide concrete targets for future experiments.

In SM, neutrinos are massless because they are purely chiral. Concerning neutrinos, we have two outstanding questions: (A) Now that we have observed neutrino oscillations that indicate neutrinos have masses, why is only one chirality (left-handed) observed? (B) Where did the neutrino mass come from? Currently, there are two major approaches to these questions: (1) Dirac mass that calls for a right-handed neutrino $\nu_R$ that is either sterile and/or super-massive, or (2) Majorana mass that requires a neutrino to be its own antiparticle.

In HM, Dirac mass doesn't work because the neutrino has only one chirality. All neutrinos are structurally left-handed, and all anti-neutrinos are right-handed. Majorana mass doesn't work either because neutrino is not identical to its anti-particle due to its substructure. So, there is no Lorentz scalar that works for a conventional mass term for neutrinos. In our view, this lack of conventional mass term naturally opens up the possibility for mixing mass terms.

As we have seen in the discussion of Figures 1 & 2 of Section 4, neutrinos with different flavors have different internal binding energies. So, we have a peculiar situation in which neutrino doesn't have a definite conventional mass state but have stable binding energies for different flavors. These stable internal binding energies exist in the framework of Weyl spinors that are Lorentz invariant. Therefore, they work perfectly as effective masses. The total energy of a neutrino is

$$E(p)^2 = p^2 + m_{eff}^2. \tag{21}$$

For a two-flavor situation with the same momentum in ultra-relativistic limit, we get the phase change after the distance L:

$$\Delta\phi = \Delta E \cdot L \simeq \frac{\Delta m_{eff}^2 \cdot L}{2E} \tag{22}$$

Using the standard mixing procedure, we get the probability of a neutrino changing its flavors:

$$P_{\alpha\to\beta}(L,E) \sim \sin^2(2\theta)\sin^2\left(\frac{\Delta m_{eff}^2 \cdot L}{4E}\right) \tag{23}$$

Where $\alpha$, $\beta$ are flavor indices and $\theta$ is a typical mixing angle. This resembles the standard neutrino oscillation with the same L/E scaling.

**Neutrinos and EDM**

In HM, it is plausible that neutrinos possess a small electric dipole moment (EDM). An EDM corresponds to a spatial separation of charges measured in the particle's rest frame. For a strictly massless particle, however, no rest frame exists, and the EDM of a massless Weyl spinor is therefore forbidden. Within HM, the possibility of a nonzero EDM thus provides an additional indication that neutrinos acquire effective masses through their substructure. Detection of even a tiny neutrino EDM would therefore constitute a signature, both confirming neutrino mass and offering a potential window into the deeper structure proposed by the model.

**Experimental Consequences**

In HM, neutrinos acquire effective masses from their internal substructure, which allows for oscillations without invoking additional mechanisms. Importantly, HM does not accommodate sterile neutrinos or Majorana states. Consequently, the model predicts the absence of right-handed (sterile or ultra-heavy) neutrinos and rules out the possibility of observing neutrino-less double beta decay ($0\nu\beta\beta$).

This perspective is not only consistent with all current experimental data but also extends beyond the Standard Model. It provides explanations for: (1) why no direct neutrino mass eigenstate has been detected, (2) why neutrinos undergo flavor oscillations while remaining purely left-chiral, and (3) why neither sterile neutrinos nor Majorana behavior have been observed.

## 8. Summary & Outlook

The hexon model offers a simple yet powerful framework in which elementary particles emerge from Weyl spinors with proto-charges. It reproduces the known fermions and bosons, predicts new dark matter candidates, and explains neutrino properties without invoking sterile states, Majorana mass terms, or Yukawa couplings. Masses, charges, chiralities, flavors and families arise naturally from the substructure of Weyl spinors with proto-charges.

Several concrete predictions distinguish the model from other Beyond the Standard Models: 1. **No proton decay**: the substructures of hadrons and leptons are constituently different. 2. **Left-handedness of neutrinos:** neutrinos are exclusively left-handed, acquiring effective mass from substructure; no right-handed (sterile or heavy) or Majorana neutrinos exist. 3. **Potential neutrino EDM:** a small but nonzero electric dipole moment may exist, offering an experimental window into its substructure. 4. **New dark matter candidates:** In the current energy regime, the model does not predict any other particles than what are in SM – except for a class of new massive neutral bosons as dark matter candidates.

The model is simple and economical – constructed from a single field (Weyl field) and a single type of charge that is consistent with the field, with no externally-added symmetries or mechanisms. Its explanatory power suggests that the hexon model may represent an important step toward a fuller understanding of elementary particles and their interactions.

## Acknowledgements

The author acknowledges research support from California Baptist University and thanks Ki-Pung Park and Haneul Kim for assistance with graphics.